\documentclass[letterpaper]{article} 
\usepackage{aaai2026}  
\usepackage{times}  
\usepackage{helvet}  
\usepackage{courier}  
\usepackage[hyphens]{url}  
\usepackage{graphicx} 
\usepackage{natbib}  
\usepackage{caption} 
\usepackage{verbatim} 
\usepackage{xcolor}
\usepackage{url}
\usepackage{amsmath}
\usepackage{amssymb}
\usepackage{amsfonts}
\usepackage{booktabs}
\newcommand{\blueurl}[1]{\urlstyle{same}\color{blue}\url{#1}\color{black}\urlstyle{tt}}
\AtBeginDocument{%
  \providecommand\BibTeX{{%
    \normalfont B\kern-0.5em{\scshape i\kern-0.25em b}\kern-0.8em\TeX}}}
\usepackage{algorithm}
\usepackage{algorithmic}
\usepackage{dsfont}

\usepackage{newfloat}
\usepackage{listings}
\DeclareCaptionStyle{ruled}{labelfont=normalfont,labelsep=colon,strut=off} 
\floatstyle{ruled}
\newfloat{listing}{tb}{lst}{}
\floatname{listing}{Listing}
\title{
Human-in-the-Loop Bandwidth Estimation for Quality of Experience Optimization in Real-Time Video Communication}
\author{
    Sami Khairy, Gabriel Mittag, Vishak Gopal, Ross Cutler 
}
\affiliations{
    \textsuperscript{\rm}Microsoft\\

}

\begin{document}

\maketitle

\begin{abstract}
The quality of experience (QoE) delivered by video conferencing systems is significantly influenced by accurately estimating the time-varying available bandwidth between the sender and receiver. Bandwidth estimation for real-time communications remains an open challenge due to rapidly evolving network architectures, increasingly complex protocol stacks, and the difficulty of defining QoE metrics that reliably improve user experience. In this work, we propose a deployed, human-in-the-loop, data-driven framework for bandwidth estimation to address these challenges. Our approach begins with training objective QoE reward models derived from subjective user evaluations to measure audio and video quality in real-time video conferencing systems. Subsequently, we collect roughly $1$M network traces with objective QoE rewards from real-world Microsoft Teams calls to curate a bandwidth estimation training dataset. We then introduce a novel distributional offline reinforcement learning (RL) algorithm to train a neural-network-based bandwidth estimator aimed at improving QoE for users. Our real-world A/B test demonstrates that the proposed approach reduces the subjective poor call ratio by $11.41\%$ compared to the baseline bandwidth estimator. Furthermore, the proposed offline RL algorithm is benchmarked on D4RL tasks to demonstrate its generalization beyond bandwidth estimation. 
\end{abstract}


\section{Introduction}

\begin{figure}[t]
  \centering
  \includegraphics[width=\columnwidth]{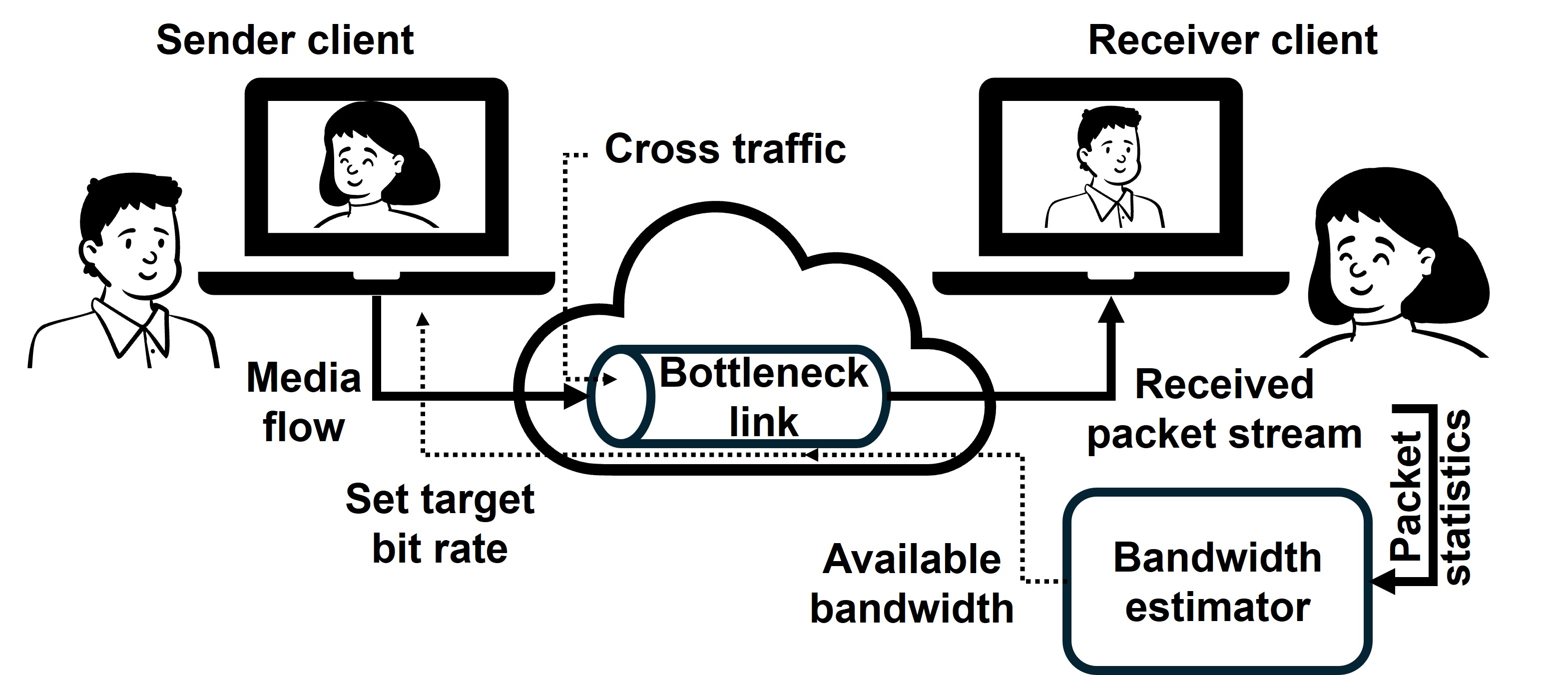}
  \caption{Bandwidth estimation in RTC: two endpoints exchange media over a time-varying bottleneck link. Cross-traffic reduces the available bandwidth. The reciever client infers available bandwidth from packet statistics (rate, delay, jitter, loss), which is fed back to the sender client to set encoder target bitrates. Accurate estimation is critical for optimising QoE, as it directly impacts video smoothness, audio clarity, and overall user satisfaction.}
  \label{fig:bwe-overview}
\end{figure}

By transforming how people connect, collaborate, and communicate across physical barriers and geographical divides, video conferencing systems have become vital for maintaining global business operations and providing accessible education \cite{markudova2023recoco,eo2022opennetlab}. The quality of experience (QoE) offered by these systems, which is a measure of a user's overall satisfaction with a multimedia conferencing system, is partly dependent on the estimation of the available bandwidth, which is defined as the bottleneck link’s unused capacity between a sender and receiver that varies over time due to fluctuations in concurrent traffic \cite{strauss2003measurement}. As illustrated in Figure \ref{fig:bwe-overview}, the receiver client estimates the available bandwidth from packet-level statistics and feeds this information back to the sender. In real-time communication (RTC) systems, this estimate guides the selection of target bitrates for the audio and video encoders, thereby regulating the sender’s transmission rate \cite{li2022reinforcement,wang2021hybrid}. Overestimating the available bandwidth results in network congestion, as the client transmits data at a rate higher than the network can handle \cite{zhang2020onrl}. Network congestion is characterized by increased delays in packet delivery, jitter, and potential packet losses, which manifest for remote meeting users as frequent resolution changes, video freezes, garbled speech, and audio/video desynchronization \cite{bentaleb2022bob, zhang2020onrl}. Conversely, underestimating the available bandwidth causes the client to encode and transmit audio/video streams at a lower rate than the network can actually support, leading to under-utilization and sub-optimal QoE. Accurately estimating available bandwidth is therefore crucial for delivering optimal QoE to users in RTC systems.

In practice, however, estimating the available bandwidth presents numerous technical and design challenges. Firstly, because of routing decisions, traffic policing, and traffic shaping mechanisms implemented by internet service providers and cloud infrastructure, network paths between senders and receivers in video conferencing systems are highly dynamic and carry fluctuating traffic loads. Secondly, there are various first and last-mile networking technologies, such as cellular (3G, 4G, 5G), Wi-Fi, WiMAX, and wired connections, each with distinct packet transmission characteristics, further complicating the estimation task. Thirdly, traffic from different applications sharing the same bottleneck link often uses different transport protocols with varying fairness mechanisms and competition dynamics, adding another layer of complexity. Together, these factors create a partially observable environment for a video conferencing client, which can only access local packet statistics to infer available bandwidth \cite{acmmmssys2024challenge}. While all estimators must be designed with these constraints in mind, evaluating and improving them requires defining end-to-end (E2E) metrics that truly capture QoE, rather than relying solely on traditional Quality of Service (QoS) metrics such as throughput, delay, and packet loss, whose correlation with QoE is context-dependent and not well understood \cite{acmmmssys2024challenge}. By focusing on E2E QoE metrics, we ensure system enhancements align with user-perceived quality, leading to more meaningful improvements. This motivates a data-driven approach that learns directly from user-aligned signals rather than proxying through QoS alone.

In this work, we propose a holistic data-driven framework for designing next-generation available bandwidth estimators suitable for real-world deployment. Specifically, our contributions are as follows.

\begin{enumerate}
    \item First, we train objective QoE reward models, which measure E2E audio and video quality. These models predict mean audio and video quality scores based on subjective user evaluations following ITU-T P.808 and P.910 guidelines.
    \item Second, we curate a comprehensive training dataset by collecting roughly $1$M network traces annotated with QoE rewards from Microsoft Teams calls. In these calls, clients used a deployed unscented Kalman filter (UKF) for baseline bandwidth estimation. 
    \item Third, we develop a novel  distributional offline reinforcement learning (RL) algorithm to train a neural network–based bandwidth estimator optimized for QoE. The proposed algorithm extends the state-of-the-art Implicit Q-learning (IQL) algorithm \cite{kostrikov2021offline} to the distributional RL paradigm to improve robustness, and employs asymmetric learning signals for the actor and critic based on domain knowledge. 
    \item Finally, we conduct extensive testbed and real-world evaluations, demonstrating significant improvements in objective QoE metrics and subjective ratings in large-scale A/B tests in production. Specifically, it is shown that  the proposed approach reduces the subjective poor call ratio by $11.41\%$ compared to the baseline estimator.
    \item In addition, to assess generalization beyond the available bandwidth estimation domain, we benchmark the proposed offline RL algorithm on standard continuous control tasks from the D4RL benchmark suite \cite{fu2020d4rl}, showing competitive performance with state-of-the-art methods.
\end{enumerate}

This work is deployed in production within the Microsoft Teams real-time media stack, serving millions of daily active users across diverse network conditions and device classes. By combining human-in-the-loop QoE modeling with offline RL, our approach closes the gap between offline policy optimization and safe large-scale deployment in latency-sensitive systems. Beyond RTC, the methodology can generalize to other networked multimedia applications where real-time resource allocation is critical. The next section reviews prior work on bandwidth estimation, QoE-driven optimization, and offline RL in networking.

\section{Related Work}

\subsection{Bandwidth estimation in RTC}

In RTC systems, available bandwidth refers to the bottleneck link capacity between a sender and a receiver, minus traffic from competing flows. This quantity fluctuates dynamically due to cross-traffic variations, routing changes, and link-layer dynamics. Accurate bandwidth estimation is critical because it drives the audio/video encoder's target bitrates. Overestimation leads to congestion and packet loss, while underestimation wastes capacity and reduces perceptual quality \cite{bentaleb2022bob,zhang2020onrl}.

Traditional bandwidth estimation schemes such as GCC \cite{carlucci2016gcc}, NADA \cite{rfc8698}, and SCReAM \cite{screamv2} are built on fixed heuristics or model-based filters that react to network-level indicators like packet delay, loss, or throughput trends. While these methods are widely deployed, they are often tuned for conservative performance and can underutilize capacity in variable conditions. Their reliance on QoS metrics as optimization targets is a key limitation: QoS does not always align with user-perceived QoE \cite{hamed2020perceptual}. QoE is influenced by complex interactions between network behavior, codec adaptation, and human perception. As a result, estimators optimized for QoS may fail to maximize actual user satisfaction, particularly in heterogeneous environments with diverse access technologies. This misalignment motivates estimators that are trained and evaluated using QoE-aligned objectives rather than QoS proxies.

\subsection{Online RL for bandwidth estimation}

RL enables agents to learn control policies by interacting with an environment to maximize cumulative rewards. Widely used continuous-control algorithms include Proximal Policy Optimization (PPO) \cite{schulman2017ppo}, Deep Deterministic Policy Gradient (DDPG) \cite{lillicrap2015ddpg}, and Soft Actor–Critic (SAC) \cite{haarnoja2018sac}. These methods have been explored for adaptive rate control in RTC as a way to replace or augment rule-based controllers. For example, R3Net \cite{fang2019r3net} trained an RL agent to adjust sending rates from packet statistics, and BoB \cite{bentaleb2022bob} integrated RL fine-tuning into an existing estimator. 

However, online RL requires extensive exploration during training, which is unsafe in production because suboptimal actions can harm user experience. To avoid this, training is often conducted in network simulators or emulators, including our prior work \cite{gottipati2023offline} and other studies that remain confined to controlled environments \cite{fang2019r3net,bentaleb2022bob}. While such platforms enable repeatable experiments, they cannot capture the full diversity of real-world cross-traffic, devices, and access networks. This mismatch, often referred to as the simulation-to-reality gap, limits the transferability of policies trained purely in artificial environments. A notable exception is OnRL, which demonstrated end-to-end online policy learning with a QoS-based reward for mobile video telephony under tightly controlled safeguards and a constrained action space \cite{zhang2020onrl}. While this establishes feasibility in a focused domain, sustaining exploration at enterprise RTC scale across heterogeneous networks and devices remains risky. By contrast, our framework trains QoE-aligned reward models from subjective studies and uses them directly for offline policy optimization on large-scale real-world logs.

\subsection{Offline RL for bandwidth estimation}

Offline RL removes the need for live exploration by learning policies from pre-collected datasets. This makes it appealing for bandwidth estimation in RTC, where safety and predictability are paramount. Nevertheless, offline RL faces challenges such as distribution shift, {in which} learned policies select actions that are not observed in the dataset, and overestimation bias for out-of-distribution actions. Algorithms like Conservative Q-Learning (CQL) \cite{kostrikov2021offline}, IQL \cite{fujimoto2021minimal}, and advantage-weighted approaches \cite{peng2019awr,nair2020awac,mah2020qwr} address these issues.

For bandwidth estimation, Mowgli \cite{agarwal2025mowgli} trained rate-control policies from passive telemetry, outperforming GCC without online training. The ACM MMSys’24 offline RL for bandwidth estimation grand challenge \cite{acmmmssys2024challenge} released a dataset collected in a {controlled testbed and emulation} environment with QoE-derived reward labels, enabling training of human-aligned offline bandwidth estimation policies but not yet reflecting large-scale production diversity \cite{pioneer, tentms, fastandfurious, gottipati2024balancing}. {The present paper advances this trajectory by pairing QoE-aligned rewards with a  distributional RL agent trained on large-scale Teams telemetry and by validating the learned policy in production A/B tests.}
\subsection{Motivation and gaps}

Building on the limitations of rule-based controllers in heterogeneous networks, the risks of online training, and the promise of offline learning, we distill four deployment-driven challenges:
\begin{itemize}
    \item \textbf{Dynamic and heterogeneous networks:} network conditions vary widely across users and over time, making it difficult for static heuristics to perform well universally.
    \item \textbf{Partial observability:} the true state of the network is not directly measurable; estimators must infer it from noisy and delayed observations.
    \item \textbf{QoE alignment:} traditional metrics like throughput and packet loss do not always correlate with user satisfaction. Estimators must optimize for perceptual quality metrics which correlates with mean opinion scores (MOS).
    \item \textbf{Safety and deployability:} online RL poses risks to user experience during training. Offline RL mitigates this by learning from historical data, enabling safe deployment.
\end{itemize}

Our design addresses these challenges in an integrated way. To cover network diversity, we train on large-scale Microsoft Teams telemetry. To mitigate partial observability, we construct compact, history-aware state from local packet timing, loss, and jitter that captures network dynamics. To align optimization with user experience, we train audio and video QoE models on subjective evaluations conducted under ITU-T P.808/P.910 and use their predictions as rewards. To preserve safety in deployment, we learn policies offline and A/B test before broad exposure.

We implement this blueprint in a deployed system that couples a human-in-the-loop data pipeline with a distributional RL agent trained offline and executed within the media stack at millisecond-scale latency. The next section details the system and learning algorithm, followed by testbed and production evaluations.

\begin{figure*}[t]
  \centering
  \includegraphics[width=\textwidth]{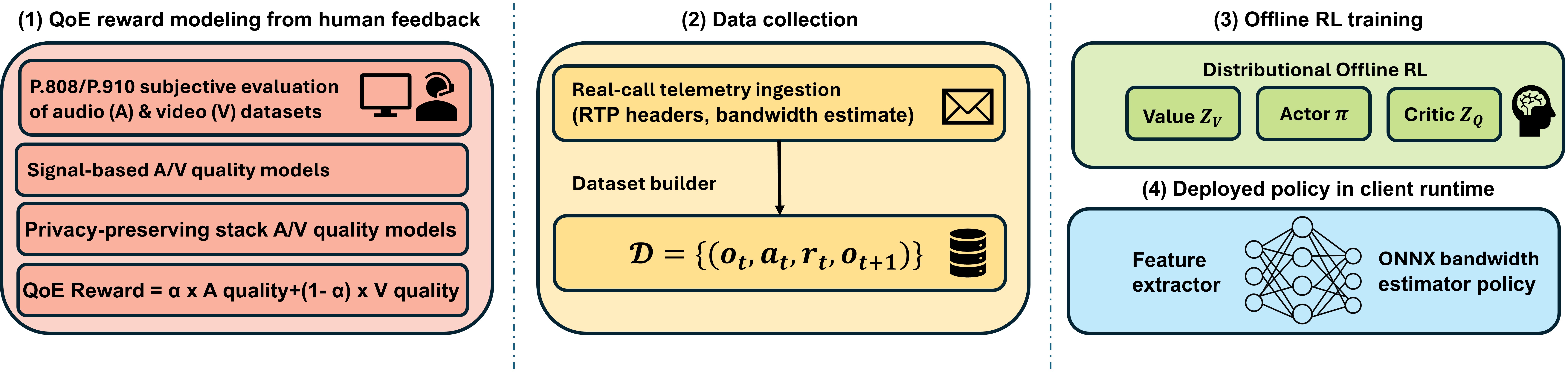}
  \caption{Overall framework: (1) QoE reward modeling from human feedback (P.808/P.910) produces in-stack audio/video quality predictors for use as a QoE reward; (2) real-call telemetry (RTP headers, baseline bandwidth estimates) is transformed into $(o_t,a_t,r_t,o_{t+1})$ trajectories; (3) a distributional offline RL agent (value $Z_V$, critic $Z_Q$, policy $\pi$) is trained and exported to ONNX for client-side inference; (4) telemetry and A/B testing close the deployment loop.}
\end{figure*}

\section{A Data-Driven Framework for Human-Aligned Bandwidth Estimation}

\subsection{QoE reward modeling from human feedback}

The use of proper reward functions that align with a user's experience of audio and video quality is crucial when training and evaluating bandwidth estimation models. Reward functions that accurately reflect user experience ensure that the models prioritize the aspects of quality that matter most to users, such as clarity, smoothness, and minimal latency. This alignment is essential because it directly impacts user satisfaction and engagement. For instance, a model that optimizes for technical metrics without considering user experience may result in high bandwidth usage without a corresponding improvement in perceived quality. Therefore, incorporating user-centric reward functions helps in developing models that not only perform well in technical evaluations but also enhance the overall user experience.

\subsubsection{\textbf{Audio quality model}}

A signal-based audio quality model was initially trained to predict the quality of received audio signals in peer-to-peer (P2P) calls. Specifically, P2P calls were conducted between pairs of machines connected through networking emulation software that emulated various network characteristics, such as burst loss, traffic policing, and bandwidth fluctuations. Due to these network transmission characteristics, the received audio signals were often distorted and corrupted with random noise, the nature of which depended on the emulated network scenario. The dataset of received audio recordings was rated using subjective scores according to ITU-T Recommendation P.808. This process involved human raters listening to the audio samples and assigning  values on a scale from $1$ to $5$, with $5$ representing the best quality. Subsequently, a no-reference Wav2Vec-based model was trained on this dataset, and achieved a high Pearson Correlation Coefficient (PCC) of $0.951$ and a Root Mean Square Error (RMSE) of $0.194$ with the subjective P.808 scores in the validation set.

While this signal-based model proved effective in laboratory experiments due to its ability to leverage detailed audio signal features, it was not practical for deployment in real-world real-time client environments due to its high complexity and the need for raw audio signals which violates user privacy. To address these issues, the signal-based model was distilled into a stack audio quality model that could run efficiently and cost-effectively in the media stack. The stack audio quality model utilizes key media metrics such as audio receive rate, jitter, and packet loss concealment, achieving a high PCC of $0.972$ with objective audio quality scores. This transition enabled real-world, real-time, privacy-preserving audio-quality evaluation without compromising accuracy.

\subsubsection{\textbf{Video quality model}} 

In a similar setup, an LSTM-based full-reference video quality model that leverages VMAF \cite{li2016toward} and freeze features to capture temporal distortions is developed \cite{mittag2023lstm}. While this model demonstrated a high PCC of $0.985$ and an RMSE of $0.202$ with subjective data provided by human raters according to ITU-T Recommendation P.910, it was also not suitable for client-side deployment due to its high complexity and the need for full video reference data. To overcome these limitations, the full-reference video quality model was distilled into a stack video quality model that relies on media metrics such as resolution, quantization parameters, the user's viewport size, freezes, and frame rate (FPS). This stack model achieved an extremely high PCC of $0.998$ with objective video quality scores, demonstrating its effectiveness in real-time applications while maintaining low computational overhead. The use of media metrics instead of raw video data ensures that the model can operate efficiently on client devices, preserving user privacy and reducing resource consumption.

\subsubsection{\textbf{QoE reward model}} 

Given the stack audio and video quality modes, we define the QoE reward as

\begin{equation} \label{r_qoe}
    r_\text{QoE} = \alpha \times \text{audio quality} + (1-\alpha) \times \text{video quality}, 
\end{equation}
where $\alpha \in [0,1]$ is a weighting parameter. This formulation allows for a balanced consideration of both audio and video quality, reflecting the overall user experience during multimedia interactions. By adjusting the parameter $\alpha$, the model can prioritize either audio or video quality based on specific application requirements or user preferences. Such a QoE reward model aligns closely with users' subjective experiences, as it integrates key perceptual metrics from both audio and video streams. Optimizing this reward model can lead to significant improvements in user satisfaction, as it ensures that both audio and video quality are maintained at optimal levels, thereby enhancing the overall multimedia experience.

\subsection{Data collection}

Training bandwidth estimation models on real-world logs offers a robust alternative to lab-emulated datasets. This approach captures detailed network traces from actual calls, enabling models to learn from realistic conditions which would be otherwise hard to emulate. These conditions include dynamic network conditions, network anomalies, user behaviors (e.g., users joining or leaving a call, switching cameras on or off, and talking or listening), and different machine hardware/software specification. In our production system, a rule-based estimator based on an Unscented Kalman Filter (UKF) is deployed. Similar to WebRTC~\cite{bergkvist2012webrtc}, UKF models network delay dynamics and adapts bandwidth estimates using static functions derived from state variables such as one-way delays, delay gradients, and loss rates. For example, it scales estimates in response to changes in delay. While extensively validated in production, the UKF’s reliance on predefined heuristics limits its adaptability to evolving network conditions, making it a strong baseline but not a complete solution. The collected logs include RTP packet headers~\cite{schulzrinne2003rtp}, UKF bandwidth estimates, and predicted audio/video quality scores by the in-stack models. These logs are transformed into trajectories containing observations, actions, and rewards suitable for offline RL. In total, we have collected approximately $1$M traces, which yield about $1.25$ trillion state-action-reward pairs in the RL setting.


\subsection{Bandwidth estimation with offline RL}

Estimating available bandwidth in real-time video conferencing is inherently challenging due to the unobservable nature of the bottleneck link between a sender and a receiver. The agent must infer bandwidth from noisy and stochastic signals derived from the received packet stream, which are influenced by cross-traffic, queueing dynamics, and other network uncertainties. We formulate bandwidth estimation as a Partially Observable Markov Decision Process (POMDP) and propose a novel training algorithm tailored to real-time video conferencing scenarios.

\subsubsection{\textbf{Bandwidth estimation as a POMDP}}

\begin{itemize}
    \item \textbf{State space}: the underlying network state includes link capacity, cross-traffic load, propagation delay, packet loss, and jitter. These factors are influenced by external conditions such as mobility, signal interference, and transport technology (e.g., 5G, satellite), making the true state unobservable and dynamic. 
    
    \item \textbf{Observation space}: observations are computed from received RTP packet headers over both short-term ($60m$s) and long-term ($600m$s) monitoring intervals (MIs). At each time step, the observation vector aggregates nine key network features: receiving rate, one-way delay, packet loss ratio, packet jitter, probabilities of video, audio, screen share, and probing packets, as well as the latest probing bandwidth estimate across 3 short-term and 3 long-term MIs. This design captures both immediate and longer-term network behaviors, providing a comprehensive, multi-scale view of network performance for robust bandwidth estimation.
\item {\textbf{Action space}}: 
the agent’s action is the available bandwidth estimate in bits-per-second (bps), which is used to set the target bit rate for media encoders at each decision step. 

\item {\textbf{Reward function}}: 
rewards reflects the predicted QoE for a given state-action pair, as defined in Eq. \eqref{r_qoe}.

\end{itemize}

\subsubsection{\textbf{Asymmetric actor-critic}}
To address the challenges of partial observability and temporal dynamics, we adopt an asymmetric actor-critic architecture. The actor network uses an LSTM module to capture temporal patterns from recent observations, enabling adaptive bandwidth predictions. The critic network, implemented as an MLP, leverages stacked historical features to estimate long-term QoE returns.
This design balances responsiveness and stability: the actor focuses on immediate adaptation using recent data, while the critic benefits from broader temporal context. Empirically, we found this separation to improve training stability and policy performance in online evaluation.

\subsubsection{\textbf{Multi-modal actor-critic}}

Bandwidth estimation presents a multi-modal learning challenge. The same observation may correspond to different bandwidth conditions depending on sender behavior, device capabilities, or media type. Similarly, QoE outcomes can vary across different devices and user's viewports even under identical network conditions.
To model this complexity, we represent both the actor and critic as a mixture density network (MDN) parameterising a Gaussian mixture (GM). Specifically, each network has an output layer with $N \times 3$ neurons, where $N$ is the number of components in the mixture. Each component $i \in [N]$ in mixture model is parameterized by the component weight in the mixture $w_i$, the component mean $\mu_i$, and the component standard deviation $\sigma_i$. In this work, we set $N=3$ unless otherwise mentioned.

\subsubsection{\textbf{Optimizing QoE with distributional offline RL}} 

To effectively optimize QoE in our bandwidth estimation setting, we develop {a distributional offline RL algorithm} based on IQL \cite{kostrikov2021offline}. The proposed approach extends the standard IQL framework into the distributional RL paradigm {\cite{bellemare2017distributional}} by modeling the entire return distribution instead of just its expected value \cite{dabney2018qr,iqn2018}. By doing so, we capture the inherent uncertainty and multi\mbox{-}modal nature of network dynamics, leading to a more robust bandwidth estimation policy. Our algorithm is an asymmetric actor–critic method tailored for offline learning of bandwidth estimators: the critic learns the QoE return distributions (distributional Q and value functions), and the actor is optimized via advantage-weighted regression using different loss functions. We describe the key components of the proposed distributional IQL (DIQL) algorithm and training objectives below.


\textbf{{Distributional value Function ($V_\psi: s \rightarrow Z_V(s)$)}}:  
rather than directly taking a max over actions, the value function $V_\psi(s)$ in IQL is learned to represent an upper envelope of the Q-function at state $s$ using expectile regression.  Since we now operate over return distributions rather than scalar values, we adopt a distributional formulation inspired by \cite{bellemare2017distributional}. Specifically, the value network outputs a distribution over returns, parameterised as a GMM: 
\[
Z_V(s) \sim \sum_{i=1}^N w_i^V(s)\,\mathcal{N}(\mu_i^V(s), \sigma_i^V(s)),
\]  
where $w_i^V(s)$ are mixture weights and $(\mu_i^V(s), \sigma_i^V(s))$ denote the mean and standard deviation of the $i$-th Gaussian component. Similarly, the Q-value distribution for $(s,a)$ is modelled as:  
\[
Z_Q(s,a) \sim \sum_{j=1}^N w_j^Q(s,a)\,\mathcal{N}(\mu_j^Q(s,a), \sigma_j^Q(s,a)).
\]  
To match these distributions, we require a metric suitable for GMMs. Following \cite{delon2020wasserstein}, we consider the Mixture Wasserstein-2 ($MW_2$) distance: 
\[
\text{MW}_2^2(Z_V,Z_Q) = \min_{\lambda \in \Pi(w^V,w^Q)} \sum_{i=1}^N \sum_{j=1}^N \lambda_{ij} \, W_2^2(i,j),
\]  
where $\Pi(w^V,w^Q)$ is the set of couplings between mixture weights and $W_2^2(i,j)$ is the squared 2-Wasserstein distance between Gaussian components:
\begin{equation}
    W_2^2(i,j) = (\mu_i^V(s) - \mu_j^Q(s,a))^2 + (\sigma_i^V(s) - \sigma_j^Q(s,a))^2.
\end{equation}

Computing the optimal coupling $\lambda$ exactly is expensive. While the Sinkhorn algorithm with entropic regularisation \cite{cuturi2013sinkhorn} can approximate it efficiently, it still requires many iterations. Instead, we adopt an upper bound using the independent coupling: 
\begin{equation}
    \widehat{\text{MW}}_2^2(Z_V,Z_Q) = \sum_{i=1}^N \sum_{j=1}^N w_i^V w_j^Q \, W_2^2(i,j),
\end{equation}
{which assumes independence between components. We find that this bound to work well in practice and avoids the computational overhead of iterative optimal transport solvers.}  {Finally, to incorporate asymmetry, we modify the cost matrix by introducing an expectile-based weight $O(\tau,i,j)$ for each pair $(i,j)$ using component means as support points, yielding the asymmetric loss:}  
\begin{equation}\label{eq:expectile-dist}
    L_V = \mathbb{E}_{(s,a)\sim \mathcal{D}}\Big[\sum_{i=1}^N \sum_{j=1}^N w_i^V w_j^Q  W_2^2(i, j) O(\tau, i, j)\Big],
\end{equation}  
where  
\[
O(\tau,i,j)=\big|\tau-\mathbf{1}_{\{z<0\}}\big|,\; z=\mu_j^Q(s,a)-\mu_i^V(s).
\]
{In the single-component average return case where $(\sigma_i^V,\sigma_j^Q)\!\to\!0$, this loss reduces to the standard IQL expectile loss \cite{kostrikov2021offline}. In the general loss however, uncertainty in the value distribution can be captured through component variances. Minimising this objective pushes $Z_V(s)$ towards the $\tau$-expectile of $Z_Q(s,a)$ while incorporating both location and scale information, providing a conservative estimate of the optimal value distribution without extrapolating to unseen actions.}

\textbf{\textit{Distributional critic ($Q_\theta: (s,a)\!\rightarrow\! Z_Q(s,a)$)}}:  
The critic network predicts the distribution of cumulative discounted QoE returns for each state–action pair. {We adopt a one-step distributional Bellman target:  
\[
Z_{\text{target}}(s,a)\;\dot=\;r(s,a)+\gamma\,Z_V(s'),
\]  
{where $r(s,a)$ is the QoE reward (Eq.~\ref{r_qoe}) and addition denotes a shift of the distribution by $r(s,a)$. Since $Z_V(s')$ is a Gaussian mixture $\{w_i,\mu_i,\sigma_i\}_{i=1}^N$, the target is also a mixture: $\{w_i,\;r+\gamma\mu_i,\;\gamma\sigma_i\}_{i=1}^N$.}  {The critic $Q_\theta$ is trained to minimise the distributional TD error between its predicted distribution and $Z_{\text{target}}(s,a)$ using a symmetric component-wise squared 2-Wasserstein loss:} 
\begin{equation} \label{eq:critic-loss}
L_Q=\mathbb{E}_{(s,a,r,s')\sim\mathcal{D}}\Big[\sum_{i=1}^N\sum_{j=1}^N w_i^{\text{target}}w_j^Q\,\Omega_2^2(i,j)\Big],
\end{equation}
{where $\Omega_2^2(i,j)=(\mu_i^{\text{target}}(s,a)-\mu_j^Q)^2+(\sigma_i^{\text{target}}(s,a)-\sigma_j^Q(s,a))^2$. This mirrors the ($MW_2$)-based approach in the value function but without the asymmetry, as the critic aims to match the Bellman target rather than form an upper envelope.}  {Because $V_\psi$ represents an upper expectile of $Q$, this loss biases $Z_Q(s,a)$ towards high-return actions in the dataset, enabling policy improvement. By modelling full distributions, the critic captures multi-modality in network scenarios where similar actions can yield both high and low QoE, avoiding the pitfalls of a single expected value. In practice, we employ two critics and select the mixture with the smaller mean to mitigate overestimation.}

\textbf{\textit{Actor policy function  ($\pi_\phi: s \rightarrow Z_\pi(s)$)}}: 
the policy network which serves as the bandwidth estimator, is also parametrized as a Gaussian mixture over the continuous action space $\pi_\phi(a|s)$. The actor is trained {via} advantage weighted regression (AWR) using the mean of the learned critic and value functions. Concretely, given a state $s$ and an action $a$, the advantage function is defined as $A(s,a) = \bar{Q}_\theta(s,a)-\bar{V}_\psi(s)$, {where} $\bar{Q}_\theta(s,a) = \sum_{i=1}^N w_i^Q \mu_i^Q(s,a)$ {and} 
$\bar{V}_\psi(s) = \sum_{i=1}^N w_i^V \mu_i^V(s)$. {Here $\bar{Q}_\theta$ and $\bar{V}_\psi$ are mixture-mean expectations, consistent with the distributional parameterisation; we stop gradients through $Q$ and $V$ when updating $\pi$.}
The policy loss is, 
\begin{equation}\label{eq:actor-loss}
    L_\pi \;=\; -\,\mathbb{E}_{(s,a)\sim d} \Big[ \exp\!\Big(\beta A(s,a)\Big)\; \log \pi_\phi(a\,|\,s)\Big]~,
\end{equation}
where $\beta>0$ is a temperature parameter. Equation~\eqref{eq:actor-loss} can be interpreted as a form of soft policy improvement: it trains $\pi_\phi$ to imitate actions in the dataset, but upweights those actions that would lead to higher-than-average returns according to the current $Q$ estimate. 


\subsubsection{\textbf{Implementation details}} 

The actor network consists of an LSTM layer with $128$ neurons, followed by $5 \times 128$ dense layers with $tanh$ activations. Dropout layers with a drop rate of $0.05$ are introduced between dense layers for regularization. Output layer parameterises a 3-component GM: weights via softmax, means via \textit{tanh} (bounded to $[-1,1]$ but then scaled to bps as in Eq. 1 \cite{acmmmssys2024challenge}), and standard deviations via \textit{softplus} with a small floor for stability. The architectures for the critic and value networks are identical except for the lack of the LSTM layer.

We train the agent on the collected dataset of real call traces, which have been pre-processed into sequences of $(o_t, a_t, r_t, o_{t+1})$. Training proceeds by alternating the three updates described in the previous subsection. For each minibatch sampled from the dataset: (i) update $V_\psi$ by minimizing $L_V$ (Eq.\eqref{eq:expectile-dist}), (ii) update $\pi_\phi$ by minimizing $L_\pi$ (Eq.\eqref{eq:actor-loss}), and (iii) update $Q_{\theta}$ by minimizing $L_Q$ (Eq.\eqref{eq:critic-loss}) with targets from the current $V_\psi$.


\section{Evaluation}

We evaluate our approach across three settings: (i) large-scale A/B testing in Microsoft Teams, (ii) controlled testbed experiments with diverse network profiles and ablations, and (iii) out-of-domain continuous control tasks to validate the proposed distributional offline RL algorithm. 

\subsection{Production A/B testing}

\paragraph{Training and deployment.} We train a bandwidth estimation model based on the proposed framework using an Azure NDm A100 v4 cluster for approximately a week. The resulting model is converted to ONNX and integrated into the Microsoft Teams media stack 
with the same feature normalization and preprocessing used in offline training to ensure parity. The estimator is invoked every $60m$s with a median inference time of roughly $600\mu$s which easily fits within real-time latency budgets. 

\paragraph{Experimental design.} A staged rollout was conducted to ensure safety: we first ran the estimator in shadow mode on a small cohort, then incrementally ramped to larger user populations. The final A/B test lasted two weeks, randomizing at the call level and treating over $25$ million calls across diverse devices and network types at a global scale. For every call leg, the objective video and audio quality scores are reported from the deployed media stack models, as well as subjective poor call rate (PCR) if users submit a rating. PCR is computed over rated calls only. We report leg-wise means and relative deltas with 95\% confidence intervals. All reported differences are significant at $p<0.05$.

\begin{table}[h!]
\centering
\caption{Production A/B outcomes. Audio/Video Quality Scores ($\uparrow$ higher is better); PCR ($\downarrow$ lower is better). }
\label{tab:ab}
\resizebox{\columnwidth}{!}{%
\begin{tabular}{l||ccccc}
\toprule
\textbf{Metric} & \textbf{Treatment} & \textbf{Control} & \textbf{Delta} & \textbf{Delta\%} & \textbf{P-value} \\
\midrule
Audio Quality Score & 4.4702 & 4.4701 & 0.0001 & +0.00\% & 0.0061 \\
Video Quality Score & 4.1721 & 4.1700 & 0.0020 & +0.05\% & 7e-38 \\
PCR & 0.0160 & 0.0180 & -0.0021 & -11.41\% & 0.0224 \\
\bottomrule
\end{tabular}}
\end{table}

\paragraph{Results.} Table~\ref{tab:ab} summarizes the outcomes. The proposed estimator consistently improves objective audio and video quality scores and reduces PCR by $11.41\%$ over the baseline estimator. PCR was pre-registered as the primary user-facing Key Performance Indicator (KPI); the $11.41\%$ relative reduction is statistically significant ($p<0.05$) and reflects materially fewer subpar calls. Video quality score shows a small but statistically significant improvement of $+0.05\%$, while the audio quality difference is statistically detectable yet practically negligible.

\begin{table*}[h]
\centering
\caption{Average QoE rewards per network profile. Our proposed DIQL algorithm demonstrates robust performance across the majority of network profiles, outperforming the behavior policy as well as prior offline policy training algorithms.}
\label{tab:netem_results}
\begin{tabular}{lccccc|c}
\toprule
Network profile & Behavior & BC & IQL & TD3BC & QR & DIQL (ours) \\
\midrule
FLB 1 & 2.377 $\pm$ 0.074 & \textbf{2.403 $\pm$ 0.022} & \textbf{2.405 $\pm$ 0.031} & \textbf{2.412 $\pm$ 0.024} & \textbf{2.423 $\pm$ 0.023} & \textbf{2.416 $\pm$ 0.029} \\
FLB 2 & 2.510 $\pm$ 0.080 & 2.538 $\pm$ 0.075 & \textbf{2.659 $\pm$ 0.066} & 2.598 $\pm$ 0.060 & 2.581 $\pm$ 0.060 & \textbf{2.661 $\pm$ 0.060} \\
FLB 3 & \textbf{2.068 $\pm$ 0.075} & 2.024 $\pm$ 0.117 & 2.028 $\pm$ 0.104 & \textbf{2.067 $\pm$ 0.078} & 2.013 $\pm$ 0.094 & \textbf{2.076 $\pm$ 0.103} \\
FLB 4 & 2.418 $\pm$ 0.083 & 2.455 $\pm$ 0.066 & \textbf{2.480 $\pm$ 0.056} & \textbf{2.463 $\pm$ 0.076} & 2.455 $\pm$ 0.070 & 2.424 $\pm$ 0.064 \\
BL 8ML25 & 3.441 $\pm$ 0.407 & 3.872 $\pm$ 0.134 & 3.865 $\pm$ 0.184 & \textbf{3.946 $\pm$ 0.142} & 3.858 $\pm$ 0.176 & 3.829 $\pm$ 0.195 \\
BL 100kL10 & 1.464 $\pm$ 0.032 & 1.463 $\pm$ 0.044 & 1.472 $\pm$ 0.024 & 1.477 $\pm$ 0.023 & \textbf{1.492 $\pm$ 0.021} & 1.476 $\pm$ 0.040 \\
BL 1ML10 & 2.733 $\pm$ 0.052 & 2.750 $\pm$ 0.076 & 2.797 $\pm$ 0.058 & 2.808 $\pm$ 0.071 & 2.822 $\pm$ 0.074 & \textbf{2.858 $\pm$ 0.056} \\
BL 400kL25 & 1.760 $\pm$ 0.082 & 1.814 $\pm$ 0.113 & 1.854 $\pm$ 0.106 & 1.865 $\pm$ 0.099 & \textbf{1.887 $\pm$ 0.083} & \textbf{1.879 $\pm$ 0.107} \\
BL 100k & 1.374 $\pm$ 0.034 & 1.388 $\pm$ 0.031 & \textbf{1.390 $\pm$ 0.042} & \textbf{1.399 $\pm$ 0.027} & \textbf{1.404 $\pm$ 0.034} & \textbf{1.403 $\pm$ 0.046} \\
BL 1ML25 & 2.306 $\pm$ 0.166 & 2.359 $\pm$ 0.178 & 2.412 $\pm$ 0.207 & \textbf{2.520 $\pm$ 0.155} & 2.427 $\pm$ 0.246 & \textbf{2.521 $\pm$ 0.220} \\
BL 400kL10 & 2.143 $\pm$ 0.056 & 2.139 $\pm$ 0.039 & \textbf{2.191 $\pm$ 0.040} & 2.172 $\pm$ 0.045 & \textbf{2.199 $\pm$ 0.040} & \textbf{2.213 $\pm$ 0.039} \\
RNDL 1ML20B & 2.628 $\pm$ 0.179 & 2.844 $\pm$ 0.100 & 2.924 $\pm$ 0.066 & 2.936 $\pm$ 0.050 & 2.905 $\pm$ 0.058 & \textbf{2.982 $\pm$ 0.072} \\
RNDL 400kL20 & 2.185 $\pm$ 0.081 & 2.138 $\pm$ 0.045 & \textbf{2.221 $\pm$ 0.033} & \textbf{2.225 $\pm$ 0.039} & \textbf{2.222 $\pm$ 0.041} & \textbf{2.227 $\pm$ 0.045} \\
RNDL 400kL20B & 2.170 $\pm$ 0.076 & 2.160 $\pm$ 0.043 & 2.211 $\pm$ 0.047 & 2.209 $\pm$ 0.026 & 2.185 $\pm$ 0.078 & \textbf{2.234 $\pm$ 0.040} \\
RNDL 100kL20 & 1.443 $\pm$ 0.037 & 1.442 $\pm$ 0.056 & 1.433 $\pm$ 0.040 & \textbf{1.466 $\pm$ 0.021} & \textbf{1.453 $\pm$ 0.033} & \textbf{1.456 $\pm$ 0.038} \\
RNDL 1ML20 & 2.630 $\pm$ 0.150 & 2.845 $\pm$ 0.084 & 2.845 $\pm$ 0.239 & 2.919 $\pm$ 0.054 & 2.909 $\pm$ 0.068 & \textbf{2.993 $\pm$ 0.108} \\
4G 700k & 2.198 $\pm$ 0.057 & \textbf{2.222 $\pm$ 0.045} & \textbf{2.229 $\pm$ 0.023} & \textbf{2.238 $\pm$ 0.042} & \textbf{2.218 $\pm$ 0.056} & \textbf{2.219 $\pm$ 0.044} \\
4G 500k & 2.070 $\pm$ 0.059 & \textbf{2.092 $\pm$ 0.059} & \textbf{2.092 $\pm$ 0.049} & \textbf{2.091 $\pm$ 0.059} & \textbf{2.081 $\pm$ 0.052} & \textbf{2.085 $\pm$ 0.042} \\
4G 300k & 1.628 $\pm$ 0.025 & 1.632 $\pm$ 0.046 & \textbf{1.660 $\pm$ 0.036} & 1.629 $\pm$ 0.033 & 1.629 $\pm$ 0.039 & \textbf{1.658 $\pm$ 0.029} \\
FB 8M & \textbf{4.295 $\pm$ 0.022} & \textbf{4.293 $\pm$ 0.017} & \textbf{4.297 $\pm$ 0.021} & \textbf{4.289 $\pm$ 0.019} & \textbf{4.289 $\pm$ 0.020} & \textbf{4.304 $\pm$ 0.019} \\
FB 4M & \textbf{4.286 $\pm$ 0.029} & \textbf{4.264 $\pm$ 0.023} & \textbf{4.283 $\pm$ 0.024} & \textbf{4.265 $\pm$ 0.020} & \textbf{4.266 $\pm$ 0.024} & \textbf{4.280 $\pm$ 0.022} \\
FB 5M & \textbf{4.289 $\pm$ 0.024} & \textbf{4.284 $\pm$ 0.022} & \textbf{4.284 $\pm$ 0.026} & \textbf{4.277 $\pm$ 0.023} & \textbf{4.275 $\pm$ 0.019} & \textbf{4.284 $\pm$ 0.034} \\
FB 3M & \textbf{4.246 $\pm$ 0.027} & \textbf{4.219 $\pm$ 0.020} & \textbf{4.258 $\pm$ 0.031} & \textbf{4.242 $\pm$ 0.023} & \textbf{4.241 $\pm$ 0.029} & \textbf{4.238 $\pm$ 0.028} \\
FB 1M & 3.251 $\pm$ 0.133 & 3.188 $\pm$ 0.034 & 3.270 $\pm$ 0.033 & 3.250 $\pm$ 0.035 & 3.241 $\pm$ 0.062 & \textbf{3.372 $\pm$ 0.044} \\
FB 800k & 2.962 $\pm$ 0.041 & 2.912 $\pm$ 0.063 & \textbf{3.061 $\pm$ 0.041} & \textbf{3.064 $\pm$ 0.026} & 3.021 $\pm$ 0.051 & \textbf{3.059 $\pm$ 0.040} \\
FB 500k & 2.501 $\pm$ 0.054 & 2.516 $\pm$ 0.036 & 2.539 $\pm$ 0.022 & 2.540 $\pm$ 0.035 & 2.535 $\pm$ 0.028 & \textbf{2.578 $\pm$ 0.067} \\
FB 340k & 2.360 $\pm$ 0.087 & 2.358 $\pm$ 0.035 & \textbf{2.388 $\pm$ 0.039} & \textbf{2.381 $\pm$ 0.027} & \textbf{2.376 $\pm$ 0.030} & \textbf{2.383 $\pm$ 0.032} \\
FB 200k & 2.133 $\pm$ 0.059 & 2.082 $\pm$ 0.081 & 2.148 $\pm$ 0.074 & 2.117 $\pm$ 0.027 & 2.120 $\pm$ 0.070 & \textbf{2.235 $\pm$ 0.043} \\
FB 150k & 1.749 $\pm$ 0.046 & 1.732 $\pm$ 0.027 & \textbf{1.794 $\pm$ 0.037} & \textbf{1.788 $\pm$ 0.049} & 1.751 $\pm$ 0.046 & \textbf{1.789 $\pm$ 0.050} \\
FB 100k & 1.534 $\pm$ 0.030 & 1.527 $\pm$ 0.048 & 1.537 $\pm$ 0.034 & 1.546 $\pm$ 0.027 & \textbf{1.569 $\pm$ 0.023} & \textbf{1.563 $\pm$ 0.037} \\
\bottomrule
\end{tabular}
\end{table*}

\subsection{Testbed experiments and ablations}

To complement the production study, we evaluate the trained bandwidth estimation model in a controlled emulation platform. {Each evaluation run is a peer-to-peer video and audio call between two lab endpoints connected through a software network emulator that replays time-varying traces of capacity, delay, and loss. Profiles are deterministic and replayable.} Each model is evaluated in $15$ peer-to-peer video calls {(two legs per call)} and over $30$ network profile, including burst loss (BL), random loss (RNDL), fixed bandwidth (FB), fluctuating bandwidth (FLB), and {4G} scenarios. {We report the average objective QoE reward and standard deviation per trace across all call legs.}

\paragraph{Baselines and variants.}
We compare the proposed distributional offline RL algorithm to the following:
\begin{itemize}
    \item \textbf{Behavior policy:} deployed rule-based bandwidth estimator used to collect logs; no learning.
    \item \textbf{Behavior cloning (BC):} a neural policy trained to imitate the UKF actions with a negative log-likelihood loss.
    \item \textbf{Implicit Q-Learning (IQL) \cite{kostrikov2021offline}:} expectile-regressed value function; actor learned via advantage-weighted regression (AWR-style).
    {\item \textbf{TD3BC \cite{fujimoto2021minimalist}.} Actor maximizes $\mathbb{E}_{a\sim\pi}[Q(s,a)]$ with a behavior-cloning regularizer $\alpha\,\mathbb{E}_{(s,a)\sim\mathcal D}[\log \pi(a|s)]$. We sweep $\alpha\!\in\!\{1.0,0.1,0.01\}$ and report the best.}
    \item \textbf{Quantile-regression crtic (QR) \cite{dabney2018qr}:} {a deep quantile network with 9 quantiles $\{\!0.1,0.2,\dots,0.9\!\}$ is first trained with quantile regression to model return distribution; actor extracted via advantage-weighted regression as in IQL. {We also sweep the number of quantiles $\{3, 6, 9\}$ and report the best.}}
\end{itemize}

{For all learned baselines and our method, we hold constant the observation space, policy/value network architectures, and hyperparameters. Only the method-specific parameters (e.g., IQL expectile, $\alpha$ for TD3BC, number of quantiles for QR) differ.}

\paragraph{Training and selection.}
{Each model is trained for 300 epochs (one full pass over the training set per epoch) with ADAM optimizer \cite{kingma2015adam} with a learning rate of $3\times10^{-5}$ and a batch size of 256 trajectories. Every $5$ epochs, we compute the mean squared error (MSE) with respect to the behavior policy’s actions. The three checkpoints with the lowest MSE are evaluated online in the testbed and we report the best. {This two-stage selection reduces variance and avoids over-fitting to offline metrics.}

\paragraph{Results.}
In Table~\ref{tab:netem_results},  {methods within the top $1\%$ of the best score in a network are typeset in \textbf{bold}.} As  Table~\ref{tab:netem_results}
indicates, the proposed DIQL algorithm for training bandwidth estimation policies achieves the highest QoE across the majority of network profiles, being in the top $1\%$ for $27/30$ of profiles, with the largest improvements in lossy network conditions. Among all methods, DIQL achieves the highest average gain ($0.0848$), and the smallest minimum gain ($-0.008$), indicating its consistent improvement and minimal performance drop compared to the behavior baseline estimator. This means that improvements do not come at the cost of occasional severe regressions; guaranteeing safe deployment in real-world systems where even rare failures can significantly impact user experience.

\subsection{D4RL benchmark}
\begin{table}[h]
\centering
\caption{D4RL MuJoCo benchmark: best normalized score.}
\label{tab:d4rl_results}
\resizebox{\columnwidth}{!}{%
\begin{tabular}{lcc}
\toprule
Environment & IQL & WIQL \\
\midrule
Halfcheetah Medium Expert & \textbf{93.4 $\pm$ 0.6} & \textbf{93.3 $\pm$ 0.5} \\
Halfcheetah Medium Replay & \textbf{44.9 $\pm$ 0.2} & \textbf{44.5 $\pm$ 0.2} \\
Halfcheetah Medium & \textbf{48.0 $\pm$ 0.1} & \textbf{47.6 $\pm$ 0.1} \\
Hopper Medium Expert & \textbf{111.9 $\pm$ 0.7} & \textbf{111.6 $\pm$ 0.4} \\
Hopper Medium Replay & \textbf{99.1 $\pm$ 2.0} & 97.8 $\pm$ 1.4 \\
Hopper Medium & 64.9 $\pm$ 3.1 & \textbf{65.9 $\pm$ 4.4} \\
Walker2d Medium Expert & \textbf{112.6 $\pm$ 0.2} & \textbf{112.7 $\pm$ 0.4} \\
Walker2d Medium Replay & 82.9 $\pm$ 0.5 & \textbf{86.3 $\pm$ 3.4} \\
Walker2d Medium & 80.5 $\pm$ 0.6 & \textbf{83.0 $\pm$ 0.8} \\
\bottomrule
\end{tabular}
}
\end{table}

Finally, we evaluate the distributional offline RL algorithm itself on standard MuJoCo tasks from the D4RL benchmark \cite{fu2020d4rl}. We use three seeds per task, evaluate every 10{,}000 gradient steps, and train for 1M gradient steps with hyperparameters matching IQL. Architectures, batch size, optimizer, discount, target update rate, and state normalization are identical to IQL for parity, and actions are squashed to \([-1,1]\) with the same policy parameterization. For the distributional value/critic networks, we use $3$ components. Evaluation uses deterministic policies (mean action) over 100 episodes per checkpoint, and we report mean~$\pm$~std of the best D4RL-normalized score across seeds. This benchmark isolates the learning algorithm from the end-to-end RTC system. It can be observed based on Table~\ref{tab:d4rl_results} that the proposed algorithm matches or surpasses IQL across tasks, indicating that the distributional offline RL design is competitive beyond the RTC domain.

\section{Conclusion}

We presented a human-in-the-loop, data-driven framework for learning bandwidth estimators in RTC, combining QoE-aligned reward modeling with a distributional  offline RL algorithm. Trained on roughly $1$M Microsoft Teams call traces and deployed in the production media stack, the learned bandwidth estimator runs every $60m$s with sub-millisecond inference time and a compact memory footprint. A two-week A/B test shows an $11.41\%$ reduction in subjective poor call rate relative to the baseline estimator, alongside statistically significant gains in objective video quality scores and a negligible change in audio quality. Three elements proved decisive in practice: (i) coupling subjective protocols (P.808/P.910) with objective reward modeling so that optimization targets reflect user-perceived quality; (ii) using offline RL to learn from real data while avoiding risky online exploration; and 
(iii) exporting an ONNX model that meets latency targets, deploying it via a staged rollout, and monitoring QoE-related A/B deltas. Beyond bandwidth estimation, the learning algorithm was evaluated on the D4RL continuous-control benchmark, and showed method-level competitiveness independent of RTC integration.

\newpage
\clearpage
\bibliography{aaai2026}

\end{document}